\documentclass{jpsj-suppl-mod}
\usepackage{txfonts}
\title{Impacts of Nuclear-Physics Uncertainty in Stellar Temperatures on the s-Process Nucleosynthesis}

\author{N.~\textsc{Nishimura}$^{1, *}$, G.~\textsc{Cescutti}$^{2, *}$, R.~\textsc{Hirschi}$^{1, 3, *}$, T.~\textsc{Rauscher}$^{4, 2, *}$, J.~\textsc{Den~Hartogh}$^{1, *}$ and A.~St.~J.~\textsc{Murphy}$^{5, *}$}

\inst{
$^{1}$Astrophysics Group, Keele University, Keele ST5 5BG, UK\\
$^{2}$Centre for Astrophysical Research, University of Hertfordshire, Hatfield AL10 9AB, UK\\
$^{3}$Kavli IPMU (WPI), University of Tokyo, Kashiwa 277-8583, Japan\\
$^{4}$Department of Physics, University of Basel, 4056 Basel, Switzerland\\
$^{5}$School of Physics and Astronomy, University of Edinburgh, Edinburgh EH9 3FD, UK\\
$^{*}$BRIDGCE UK Network, www.bridgce.ac.uk, UK}

\email{n.nishimura@keele.ac.uk}

\recdate{September 10, 2016}

\abst{We evaluated the uncertainty relevant to s-process nucleosynthesis using a Monte-Carlo centred approach. We are based on a realistic and general prescription of temperature dependent uncertainty for the reactions. We considered massive stars for the weak s-process and AGB stars for the main s-process. We found that the adopted uncertainty for (n,$\gamma$) rates, tens of per cent on average, affect the production of s-process nuclei along the $\beta$-stability line, while for $\beta$-decay, for which contributions from excited states enhances the uncertainty, has the strongest impact on branching points.}

\kword{s-process nucleosynthesis, stellar evolution, nuclear reaction}

\begin{document}

\newcommand{\apj}{Astro. Phys. J. }
\newcommand{\apjs}{Astro. Phys. J. Sup. }
\newcommand{\aap}{Astronomy and Astrophysics}
\newcommand{\mnras}{Monthly Notices of the Royal Astronomical Society}

\maketitle

\section{Introduction}
The s-process nucleosynthesis is a source of heavy elements beyond iron in the universe, taking place in stellar burning environments. There are two astronomical conditions and corresponding classes of the s-process (see a review\cite{2011RvMP...83..157K} and references therein). The s-process occurs in (i) thermal pulses of low mass AGB stars producing heavy nuclei up to Pb and Bi, called the {\it main} s-process; (ii) He-core and C-shell burnings of massive stars representing lighter components up to $A \approx 90$, categorised as the {\it weak} s-process.

In both cases, the primary mechanism is to produce heavier elements due to the neutron capture and $\beta$-decay along stable isotopes from seed Fe nuclei over a long-term stellar evolution period. Neutron sources reactions for the s-process are $\alpha$-captures to different nuclei, where $^{13}{\rm C}(\alpha,{\rm n})^{16}{\rm O}$ and $^{22}{\rm Ne}(\alpha,{\rm n})^{15}{\rm Mg}$ are main reactions for the main and weak s-processes, respectively. The impact of these key fusion reactions has been studied well with focuses on several aspects \cite{2011RvMP...83..157K}. The remaining problem is that the effects of uncertainty of (n,$\gamma$) and $\beta$-decay on the final products. As a lot of these reactions involves the s-process, the uncertainty is not as simple as the cases of neutron source/poison reactions. More systematic studies based on the Monte-Carlo (MC) and statistical analysis \cite{2015JPhG...42c4007I, 2016MNRAS.463.4153R} are necessary for such problems.

In this study, we investigate the impact of uncertainty due to nuclear physics on the s-process using the MC-based nuclear reaction network \cite{2017arXiv170100489N}. Adopting simplified stellar models that reproduce typical s-process patterns, we apply realistic temperature-dependent uncertainty of nuclear reaction and decay rates to nucleosynthesis calculation. Based on an MC method, we evaluate uncertainty of nucleosynthesis yields.

\section{Methods}

We use simplified stellar evolution models in the solar metallicity based on 1D evolution calculation. We follow nucleosynthesis evolution along temporal history of the temperature and density from the initial abundances. The thermal evolution is treated as the time evolution for a ``trajectory'' as a single fluid component. We adopt $25 M_\odot$ massive star evolution model \cite{2004A&A...425..649H, 2008IAUS..255..297H} and $2M_\odot$ AGB star model calculated by the MESA code \cite{2011ApJS..192....3P}. We confirmed that these trajectories reproduce a typical abundance pattern for the main and weak s-process, respectively.

\begin{figure}[b]
\centering
\includegraphics[height=0.32\hsize]{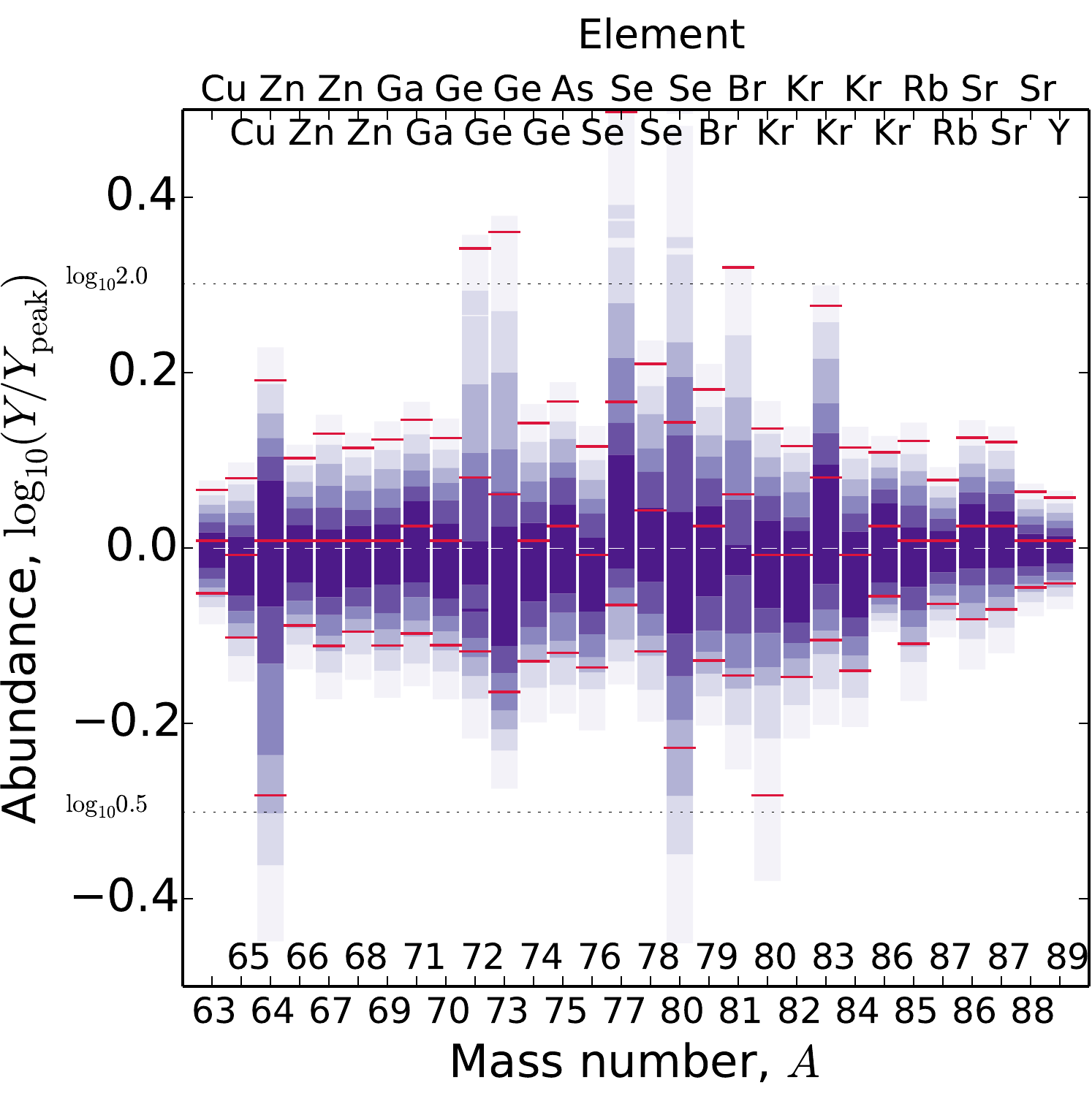}
\includegraphics[height=0.32\hsize]{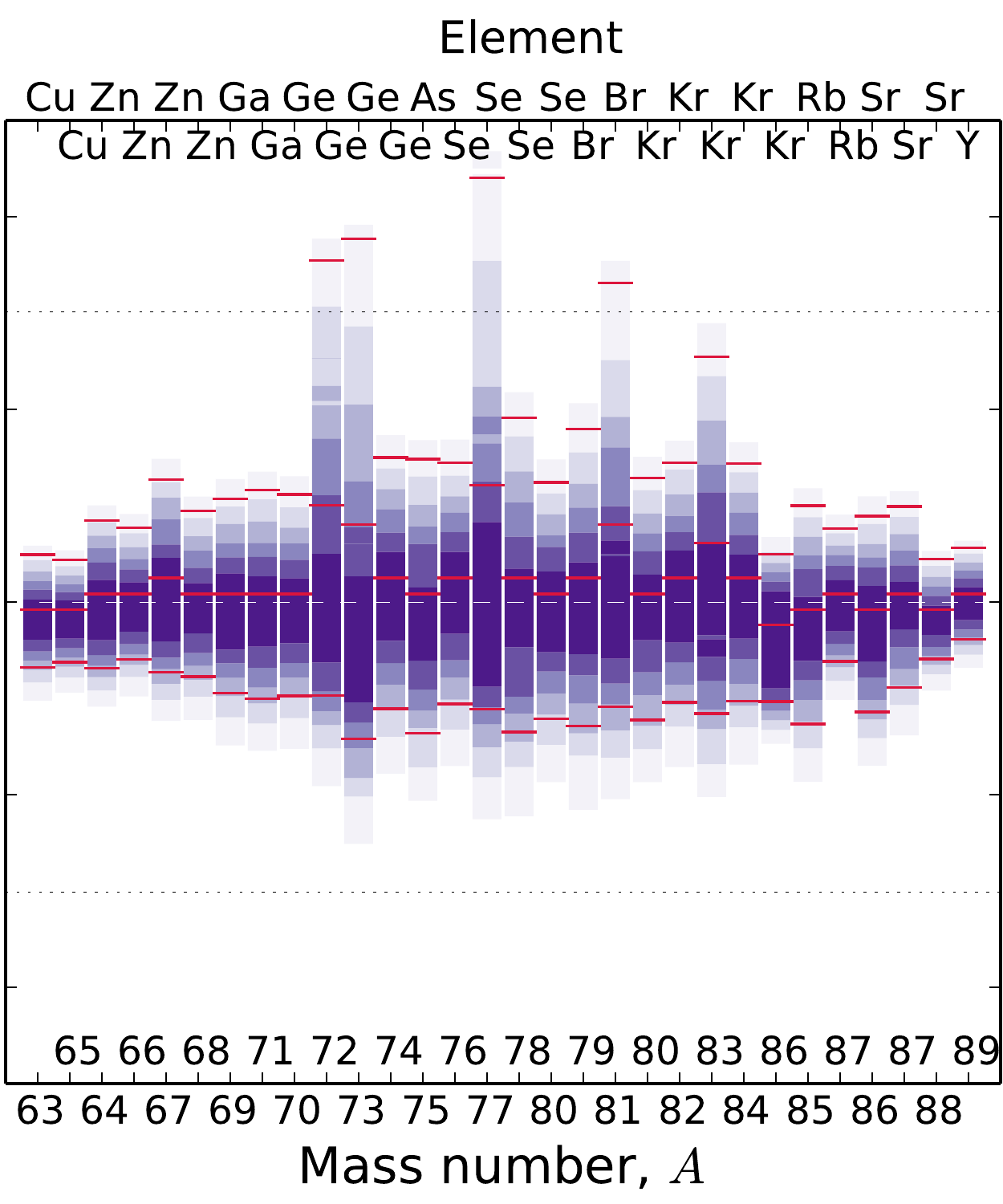}
\includegraphics[height=0.32\hsize]{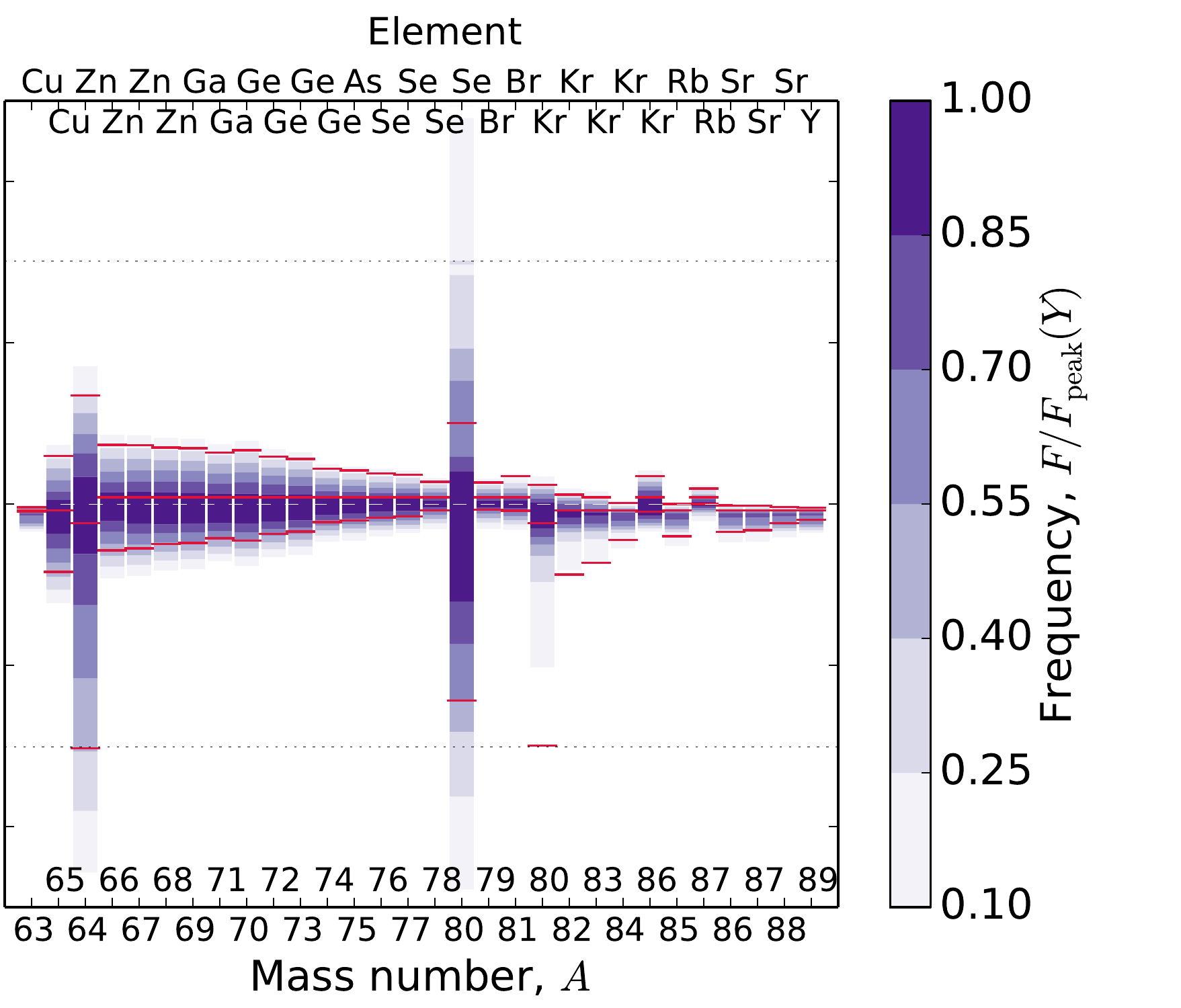}
\caption{The results of the MC for the weak s-process. Uncertainty range is shown for each isotope with red  lines covering $90$\% from the peak value for variation models of {\tt ngbt} (left), {\tt ng} (middle) and {\tt bt} (right).}
\label{fig-mc-ws}
\end{figure}

We consider that reaction rates have a temperature-dependent uncertainty due to the relative contributions by the ground state and excited states for experimental based cross sections. Following the prescription in \cite{2011ApJ...738..143R, 2012ApJS..201...26R}, experimental uncertainties are used for the ground state contributions to (n,$\gamma$) rates, whereas a factor $2$ is used for excited state uncertainties  (for details, see \cite{2012ApJS..201...26R}). As theoretical calculated rates may have large uncertainty, we simply apply a constant value $2$.

A similar approach is used for $\beta$-decay rates, based on partition functions to consider excited state contribution. The uncertainty at lower temperatures ($T < 10^7$~K) corresponds to the ground state value, while the uncertainty becomes larger as the temperature increases. We adopt a factor $5$ for the maximum value at a high temperature, although uncertainty is about $2$ in stellar burning temperatures.

\section{Results of MC calculations}
\label{sec-1}

\begin{figure}[t]
\centering
\includegraphics[height=0.29\hsize]{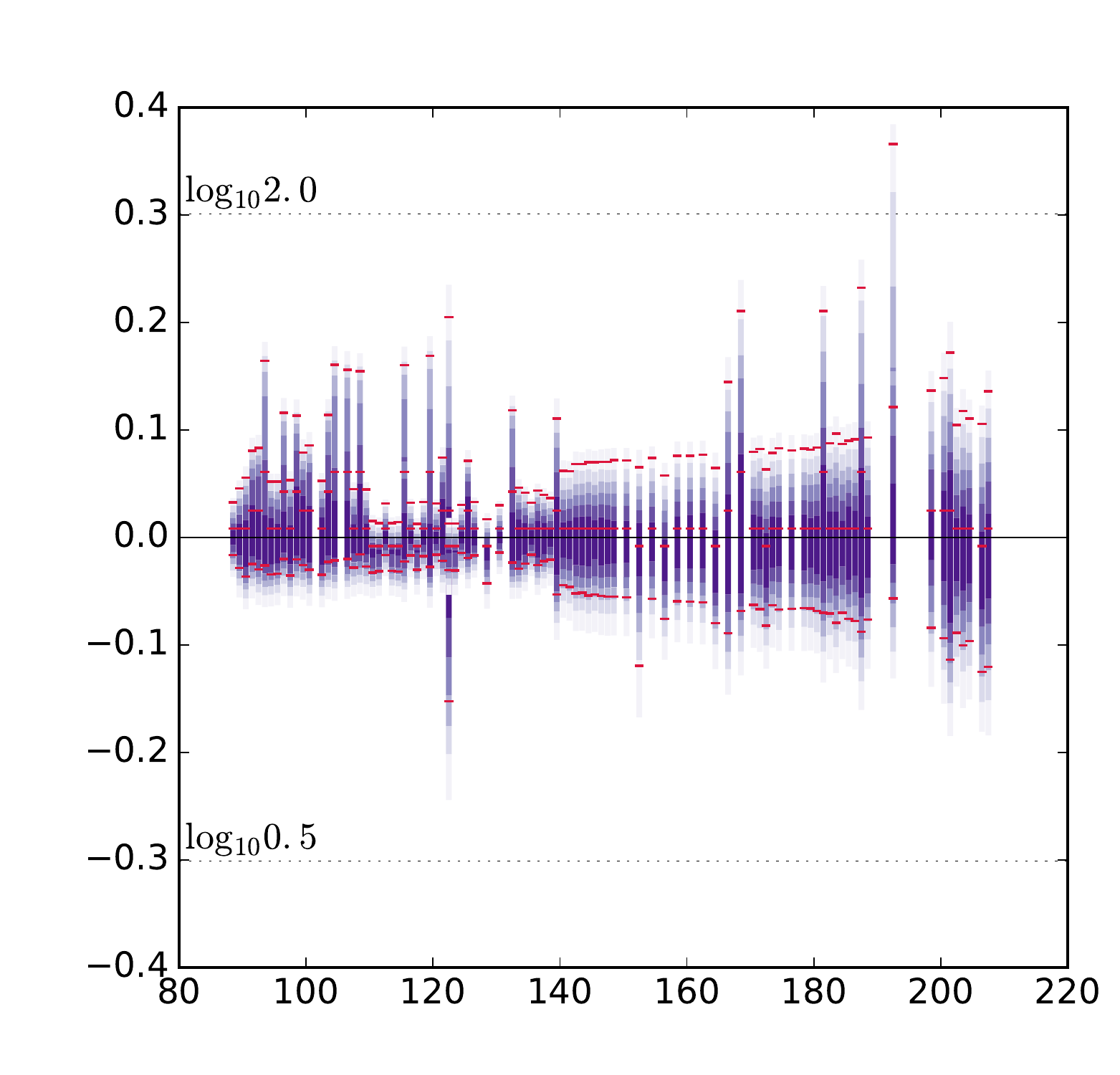}
\includegraphics[height=0.29\hsize]{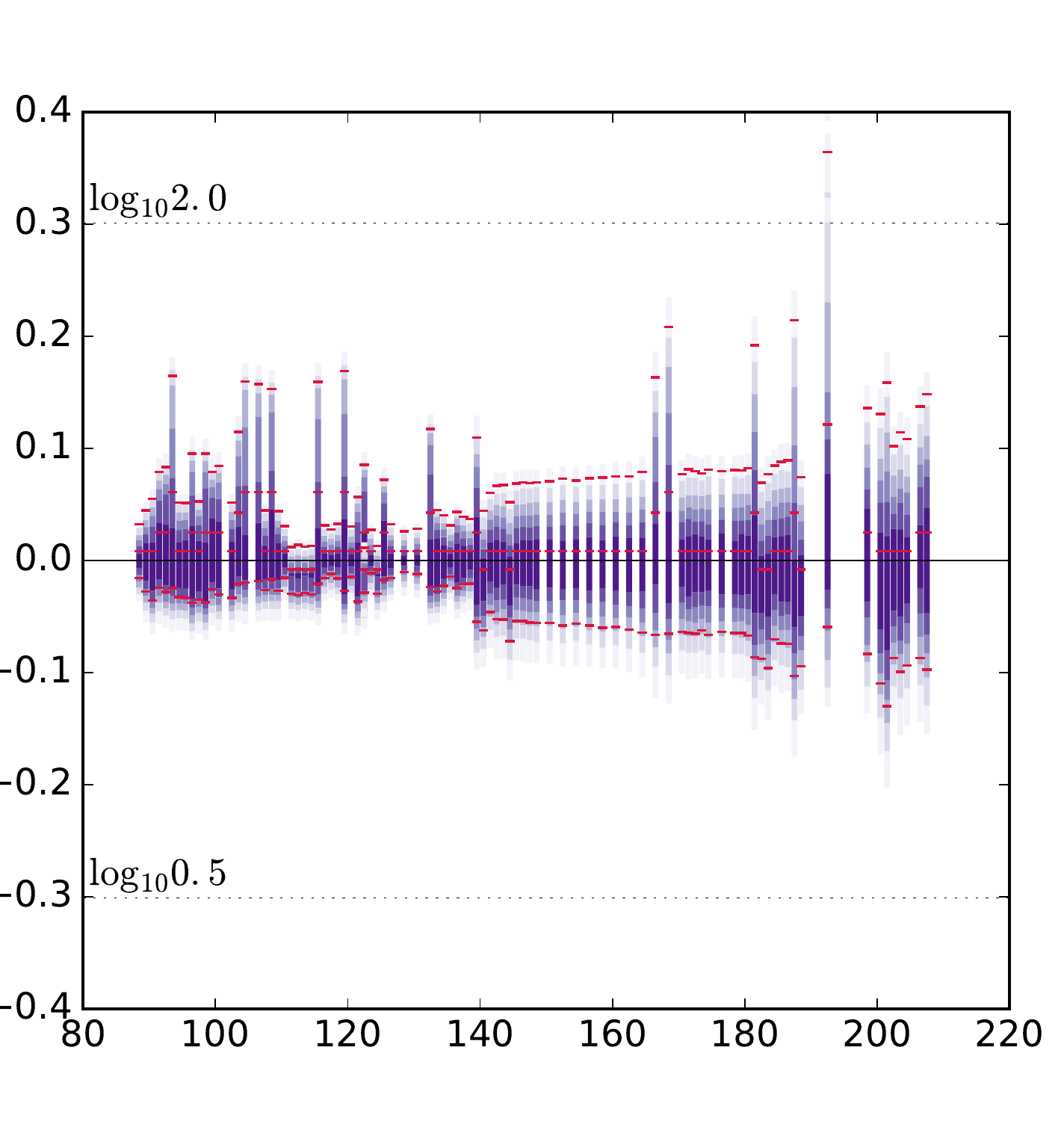}
\includegraphics[height=0.29\hsize]{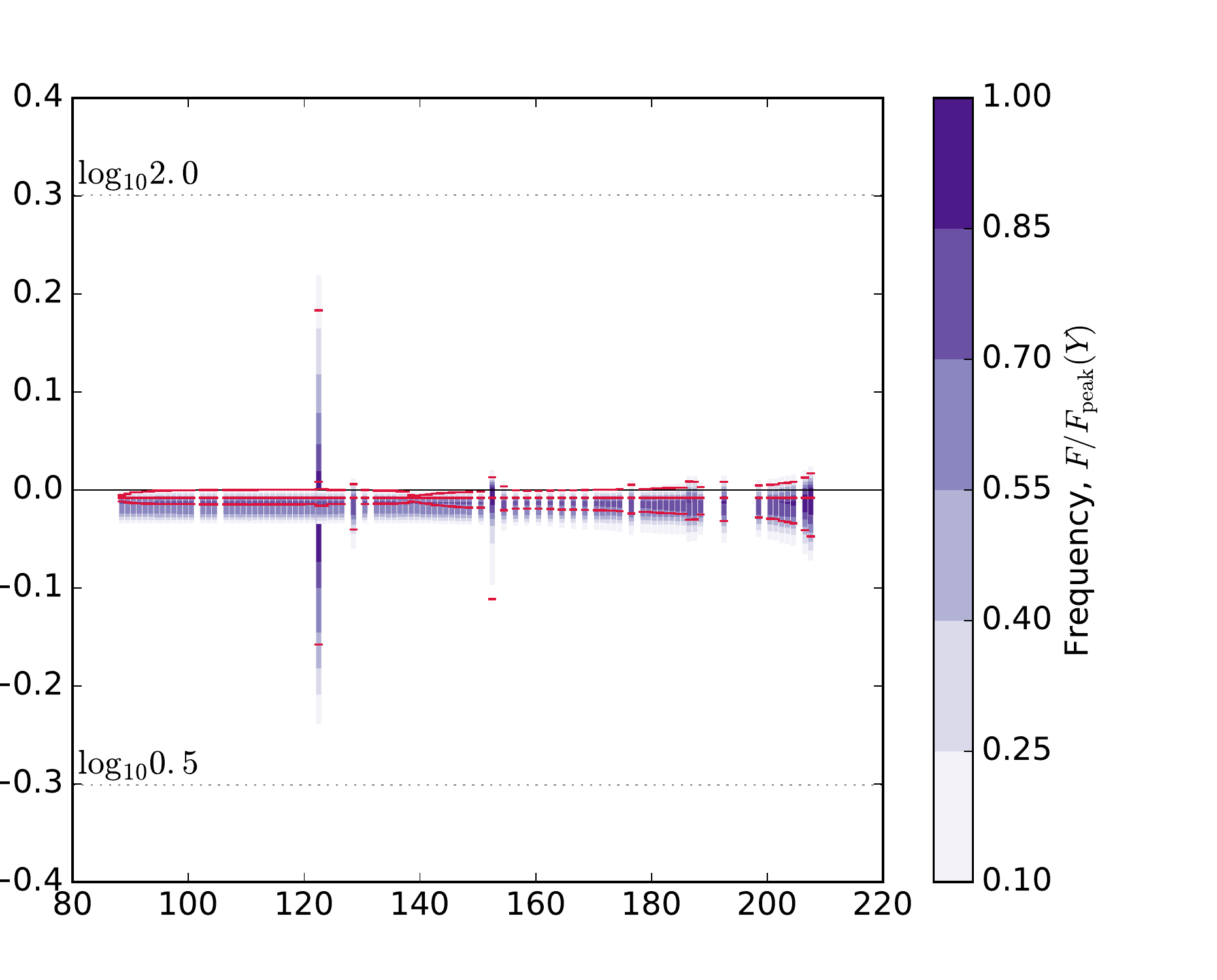}
\caption{The same as Fig.~\ref{fig-mc-ws}, but the results of main s-process for {\tt ngbt} (left), {\tt ng} (middle) and {\tt bt} (right).}
\label{fig-mc-ms}
\end{figure}

We performed MC simulations with variation of reaction rates. A uniform random distribution between the upper and lower limit of the reaction rate at a given temperature was used for each variation factor. We perform three different cases for the combination of rate variation; {\tt ngbt}: all (n,$\gamma$) and $\beta$-decay rates; {\tt ng}: only (n,$\gamma$) rates; {\tt bt}: only $\beta$-decay rates.

Fig.~\ref{fig-mc-ws} shows the resulting production uncertainty of weak s-process for the cases where we varied all (n,$\gamma$) reactions and $\beta$-decays. We select abundance uncertainties for stable s-process isotopes up to $\sim 90$. The colour distribution corresponds to the normalized probability density distribution of the uncertainty in the final abundance.

Seeing the {\tt ngbt}, $90\%$ uncertainty range of abundances within a factor of $1.5$ ($0.176$ in $\log_{10}$) region, while some isotopes show a larger uncertainty that reaches factor $2$. By the comparison of {\tt ng} and  {\tt bt}, we can understand total uncertainty is mostly due to (n,$\gamma$) reaction. Uncertainty for a few isotopes ($^{64}{\rm Zn}$ and $^{80}{\rm Se}$) affected by $\beta$-decay around branching points, although the effects of $\beta$-decay to the global isotopes are minor compared with (n,$\gamma$). This feature is also remarkable for the case of main s-process in Fig.~\ref{fig-mc-ms}. The uncertainty of $^{122}$Sn is larger due to $\beta$-decay, while most of total uncertainty is caused (n,$\gamma$) as the results of {\tt ngbt} and {\tt bt} are almost identical for majority of nuclei.

Seen in the results of {\tt bt} (in Fig.~\ref{fig-mc-ws} and \ref{fig-mc-ms}), few $\beta$-decay may cause larger uncertainty in nucleosynthesis. As we quantitatively analyse MC results calculating the correlation between decay rates and final abundances (see, \cite{2017arXiv170100489N, 2016MNRAS.463.4153R}), we find that $^{64}{\rm Cu} (\beta^{+}) ^{64}{\rm Zn}$ and $^{80}{\rm Br} (\beta^{+}) ^{80}{\rm Kr}$ have dominant impact on the production of $^{64}{\rm Zn}$ and $^{80}{\rm Se}$ for the weak s-process, respectively. Besides, $^{122}{\rm Sb} (\beta^{+}) ^{122}{\rm Te}$, competing with the $\beta^{-}$-decay counterpart, is a dominant rate for the uncertainty of $^{122}{\rm Sn}$ in the main s-process. These $\beta$-decay rates are around the s-process branching points as indicated in previous investigation.

\section{Conclusion}

We evaluated the impact on s-process nucleosynthesis in massive stars and low mass AGB stars of nuclear physics uncertainties using MC calculations. The method can identify the importance of reactions and we found that (n,$\gamma$) reactions dominate the total uncertainty, with a few important contributions from $\beta$-decays around branching points. Our method is a robust way to identify key reaction rates to support further investigations in nuclear astrophysics regarding the s-process. As we mostly focused on the effects of overall (n,$\gamma$) and $\beta$-decay in the presented study, more detailed analysis will be shown in our upcoming papers.

This project has been financially supported by the ERC (EU-FP7-ERC-2012-St Grant 306901, EU-FP7 Adv Grant GA321263-FISH) and the UK STFC (ST/M000958/1). Parts of computations were carried out by COSMOS (STFC DiRAC Facility) at DAMTP in University of Cambridge.

\end{document}